\documentclass{aa}

\usepackage[authoryear]{natbib}
\bibpunct{(}{)}{,}{a}{}{;} 
\usepackage{graphicx}
\newcommand{\msun}{{M$_\odot$}\ }
\def\msun{M$_\odot$}
\def\lsun{L$_\odot$}
\def\mhtwo{M$_{\rm H_2}$}
\def\mhi{M$_{\rm HI}$}
\def\kms{km\,s$^{-1}$}
\def\halpha{H$\alpha$}
\def\arcsec{$^{\prime\prime}$}
\def\cm{cm$^{-2}$}
\begin{document}
 
\title{Abundant molecular gas in the intergalactic medium of Stephan's Quintet}

\titlerunning{Abundant molecular gas in  
Stephan's Quintet}

\author{
Ute Lisenfeld\inst{1}
\and Jonathan~Braine\inst{2}
\and Pierre-Alain Duc\inst{3}
\and St\'ephane Leon\inst{1,4}
\and Vassilis Charmandaris\inst{5}
\and Elias Brinks\inst{6,7}
} 
\offprints{Ute Lisenfeld, ute@iaa.es}

\institute{Instituto de Astrof\'\i sica de Andaluc\'\i a, CSIC, Apdo. 3004,
 18040 Granada, Spain 
\and Observatoire de Bordeaux, UMR 5804, CNRS/INSU, B.P. 89, 
  F-33270 Floirac, France
\and CNRS URA 2052 and CEA/DSM/DAPNIA Service d'Astrophysique, Saclay,
91191 Gif sur Yvette Cedex, France
\and Physikalisches Institut, University of Cologne, Germany 
\and Cornell University, Astronomy Department, Ithaca, NY 14853, USA
\and Departamento de Astronom\'{\i}a, Universidad de Guanajuato, 
Apdo.\ Postal 144, Guanajuato, Gto 36000, Mexico
\and INAOE, Apdo. Postal 51 \& 216, Puebla, Pue 72000, Mexico
}

\date{Received  May 24 2002/ Accepted August 27 2002}

\authorrunning{Lisenfeld et al.}
 
\abstract{
Stephan's Quintet (SQ) is a system consisting of at least
four interacting galaxies  which is well known for
its complex dynamical and star formation history.
It possesses a rich intergalactic
medium (IGM), where  hydrogen clouds, both atomic
and molecular, associated with two starbursts  (refered to as
SQ~A and B) have  been found.
In order to study the extent, origin and fate of
the intergalactic molecular gas and its relation
to the formation of stars outside galaxies and
Tidal Dwarf Galaxies (TDGs), we mapped with 
the IRAM 30m antenna the carbon monoxide (CO) towards
several regions of the IGM in SQ.
In both SQ~A and B, we detected unusually large
amounts of molecular gas ($3.1 \times 10^9$ \msun \ and $7 \times 10^8$ \msun,
respectively).
In contrast, no significant CO detection was achieved
towards HII regions south of the pair NGC~7318a/b despite
their high \halpha \ luminosities. 
The molecular gas is very extended in both SQ~A and SQ~B,
over areas of between 15 and 25 kpc.
The CO clouds seem to have otherwise different
properties and may be of a different nature.
The integrated CO line of SQ~A is in particular
much wider than in SQ~B. Its CO spectrum 
shows emission  at two velocities (6000 and
 6700 \kms) that are coincident with
two HI lines. The strongest emission at 6000 \kms \
is however spatially offset from the HI emission and situated
on a ridge south-east of the starburst region. 
In SQ~B the CO emission coincides with 
that of tracers of star formation (\halpha, 15 $\mu$m and radio continuum).
The CO peak lies slightly offset from the HI peak towards
a steep HI gradient. This is indicating that the
molecular gas is forming in-situ, possibly in a region of compressed HI, 
with subsequent star formation. The star forming region
at SQ~B is the object in SQ that most resembles 
a TDG.
}
 
\maketitle
 
\section{Introduction}

\begin{figure*}
\resizebox{\hsize}{!}{\rotatebox{270}{\includegraphics{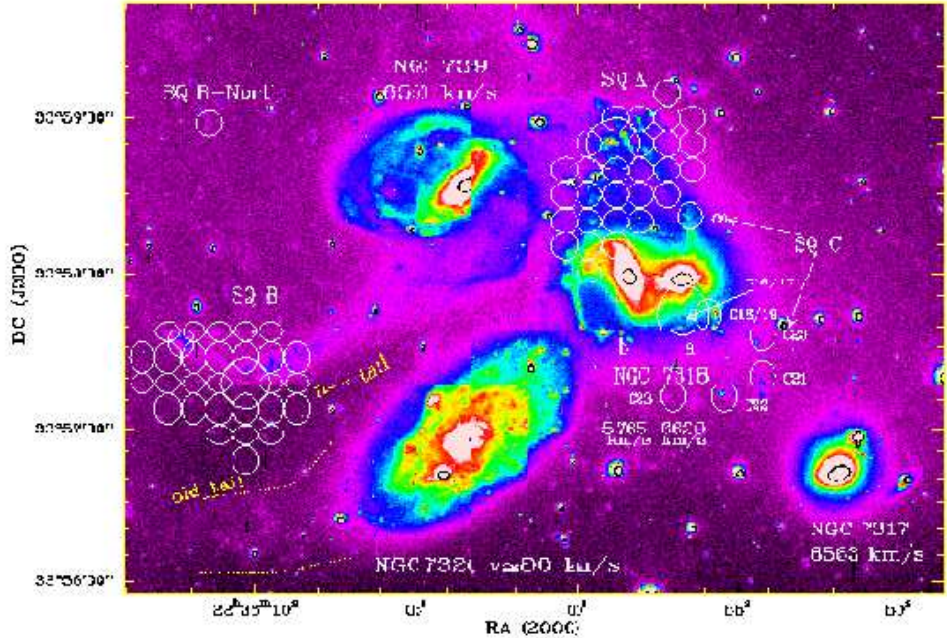}}} 
\caption{
An archival V-band image from CFHT   of Stephan's 
Quintet, showing the positions observed by us.
NGC~7320 is a foreground galaxy. The fourth
member of the group, NGC~7320c lies about 4 arcmin east of NGC~7319.
The velocities are taken from \citet{Sulentic01}.
The positions observed in CO
are indicated by circles. The large  circle
shows the central (i.e. offset 0,0) position in each region
and  gives the size of the CO(1--0) beam. 
}
\label{opt_map}
\end{figure*}

Stephan's Quintet (Hickson Compact Group 92; hereafter SQ)
is one of the best studied examples of a 
Hickson Compact Group. It 
contains four interacting galaxies, NGC~7319,
NGC~7318a, NGC~7318b, and NGC~7317. A fifth galaxy,
NGC~7320, happens to be a foreground object.  An optical image of
the group is presented in Fig.~\ref{opt_map},
\citep[see also the excellent image in][]{Arp76}.
One of its most striking properties is that 
the major part of the gas is in the intragroup medium, most
likely the result of interactions in the past and present.

A plausible scenario for the dynamical history of SQ is presented
by \citet{Moles97}.
They suggest that
a few times 10$^8$ yr ago the group experienced a collision with
NGC~7320c, a galaxy $\sim$4\,arcmin to the east of NGC~7319
but with a very similar recession velocity 
\citep[6583 \kms,][]{Sulentic01} to the other galaxies 
in SQ.
This collision removed most of the gas of NGC~7319 towards the west 
and east,  and produced the eastern
tidal tail which connects to NGC~7319.  
%
%
Presently, the group is
experiencing another collision with the ``intruder'' galaxy NGC~7318b
which strongly affects the interstellar medium (ISM) removed during
the first collision. NGC~7318b  has a recession velocity of
5765\,\kms, in contrast to the other members of the group which have
velocities close to 6600\,\kms. 
\citet{Sulentic01}, in a multi-wavelength study of the
group, confirm this scenario and suggest that the group has been
visited twice by NGC~7320c. The first collision created the very faint tidal
arm east of the interloper NGC~7320, whereas the second interaction
produced the tidal arm which stretches from NGC~7319 eastwards.

This violent dynamical history has induced star formation at various
places outside the individual galaxies.  ISOCAM mid-infrared and
H$\alpha$ observations have revealed a starburst region \citep[object
A in][hereafter called SQ~A]{Xu99} at the intersection of two faint optical
arms stemming north from NGC~7318a/b. Several knots of star
formation are visible in the tidal arm extending from NGC~7319
to the east. At this position, identified as B by \citet{Xu99} and
hereafter called SQ~B, there is also mid-infrared and H$\alpha$
emission, although much weaker than in SQ~A. 

Using HST observations, \citet{Gallagher01} found 
115 candidate star clusters, most of them distributed
among the tidal debris of  SQ.
From color-color diagrams they estimated their ages ranging from 
2--3 Myr up to several Gyr. The distribution of ages sheds light on 
the star formation history in SQ:
The youngest star clusters (with ages of less than 10 Myr) 
are found in SQ~A and south of NGC~7318a/b, while somewhat
older star clusters, with ages between 10 and 500 Myr are in the 
young tidal tail  and around NGC~7319. This is consistent 
with the picture 
that the eastern tidal tail was  produced in a previous
interaction whereas the collision and star formation around SQ~A
is ongoing.
\citet{Mendes01} have carried out Fabry-Perot observations of the
\halpha \, emission in SQ~A and south of NGC~7318a/b, from which
they determined the velocity curves for several \halpha \ emitting 
regions.  SQ~B was not covered by their observations.
They identified seven Tidal Dwarf Galaxy (TDG) candidates, selected as objects 
exhibiting a velocity gradient compatible with rotation and 
possessing an $L_{\rm B}/L_{\rm H\alpha}$ ratio consistent with that of a 
dwarf galaxy.
%

Abundant atomic hydrogen is present to the east of the 3 central
galaxies, around SQ~A and south of NGC~7318a \citep[][see their Fig. 5]
{Shostak84,Williams02}.
\citet{Williams02} suggest that the atomic gas towards the east of SQ 
consists of two tidal features, each connected to one of the 
optical tidal arms (the old and the new one, see Fig.~\ref{opt_map}).
Molecular gas has been found at the
position of SQ~A by \citet{Gao00} with BIMA and by \citet{Smith01}
with the NRAO 12m radio telescope.  Both the molecular and the atomic
gas in this location present two velocity components, 
centered at about 6000 and 
6700\,\kms, which implies that they originate from different
galaxies.  \citet{Braine01} in a study of the CO emission of 
TDGs detected 2.9$\times$10$^8$\,\msun \  of molecular gas
at SQ~B. 
In order to follow up on these detections and to elucidate the extent, 
origin and fate of the CO in the intergalactic medium (IGM) in SQ, we  
embarked on a single dish survey,
the results of which we report in this paper.

Details of our observations can be found in Section 2. In Section
3 we present the results of these observations, as well as
a comparison between the 
molecular gas and emission at other wavelengths.
We discuss our findings in Section 4 and
summarize our conclusions in Section 5. 
Following \citet{Williams02}, who assumed a recession velocity of 
6400 \kms \ and $H_0$=75\,km\,s$^{-1}$\,Mpc$^{-1}$,
we adopt a
distance of 85\,Mpc, in which case 10\arcsec \ correspond to 4.1\,kpc.

\section{Observations}       

We observed the CO(1--0) and CO(2--1) lines at 115 and 230\,GHz in SQ
in July 2001 with the IRAM 30-meter telescope on Pico Veleta.  Dual
polarization receivers were used at both frequencies with the 512
$\times$ 1 MHz filterbanks on the CO(1--0) line and the autocorrelator
or 256 $\times$ 4 MHz filterbanks on the CO(2--1).  
The observations were done in wobbler switching mode with a 
wobbler throw of 120\arcsec \ {  in azimuthal direction.}
Pointing was
monitored on nearby quasars every 60 -- 90  minutes, 
{  the rms offset being $\approx 3$\arcsec.}
At the
beginning of the observations and every time the frequency was
changed, the frequency tuning was checked by observing Sagittarius B2.
System temperatures were generally quite good, 150--200\,K at 115\,GHz
and 200--300\,K at 230\,GHz on the $T_{\rm A}^*$ scale.  The IRAM forward
efficiency, $F_{\rm eff}$, was 0.95 and 0.91 at 115 and 230\,GHz and
the beam efficiency, $B_{\rm eff}$, was 0.75 and 0.54, respectively.
Our half-power beam size was 21$^{\prime\prime}$ at 115\,GHz and
11$^{\prime\prime}$ at 230\,GHz. All CO spectra and luminosities are
presented on the main beam temperature scale ($T_{\rm mb}$) which is
defined as $T_{\rm mb} = (F_{\rm eff}/B_{\rm eff})\times T_{\rm A}^*$.
For the data reduction, the spectra were summed over the individual
positions and a constant continuum level 
was subtracted.

We observed three different regions of Stephan's Quintet: 
(i) the region around starburst SQ\,A, where strong mid-infrared
and H$\alpha$ emission had been detected {  \citep{Xu99}},   
(ii) the area around a star forming region
in the  eastern tidal arm of NGC~7319
(SQ~B), as well as  one position (SQ~B-North)
at the northern tip of the HI cloud covering SQ~B {  \citep{Williams02}}, 
and 
(iii) several positions to the south and north of NGC~7318a/b, 
observed by \citet{Mendes01} in \halpha.
We call these positions SQ~C$n$ where $n$ is the number that
\citet{Mendes01}  gave to the corresponding region.
The 
locations of our pointings are shown in Fig.~\ref{opt_map}. 

In SQ~A and
SQ~B the spacing of the individual pointings is 10$^{\prime\prime}$ so
that the CO(2--1) map is undersampled. Note that our observations
are centered at the positions called SQ~A and SQ~B as in \citet{Xu99}
but that they comprise a much larger area. 

\section{Results}

\subsection{Molecular gas distribution}

\begin{figure}
\resizebox{\hsize}{!}{\rotatebox{270}{\includegraphics{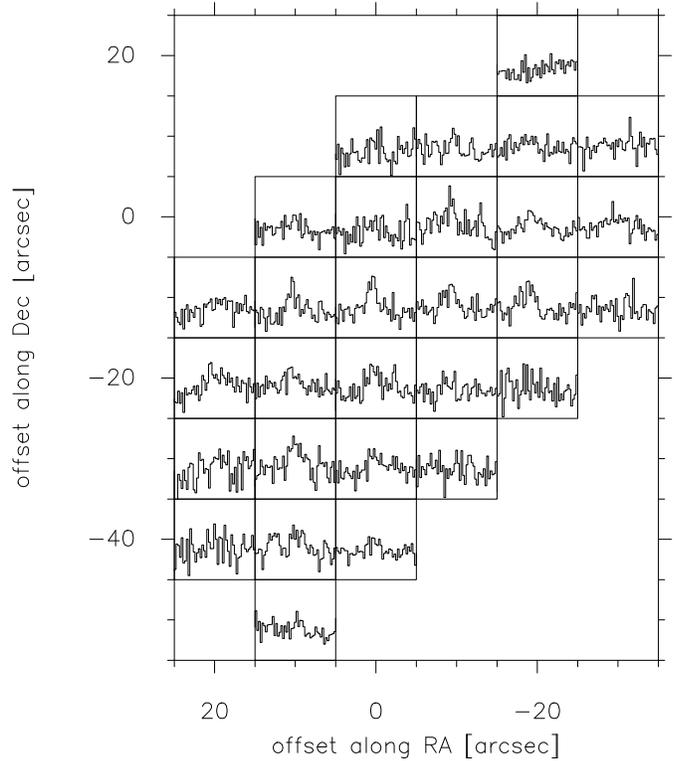}}} 
\caption{CO(1--0) emission in region A, with a velocity resolution
of 10.6\,\kms. The x-scale in the individual spectra 
is velocity and ranges from 5800 to 6300\,\kms.
The y-scale is in $T_{\rm mb}$ and ranges from -13 mK to 25 mK.
The noise is on average 3 mK.
}
\label{map_co_one_a}
\end{figure}

\begin{figure}
\resizebox{\hsize}{!}{\rotatebox{270}{\includegraphics{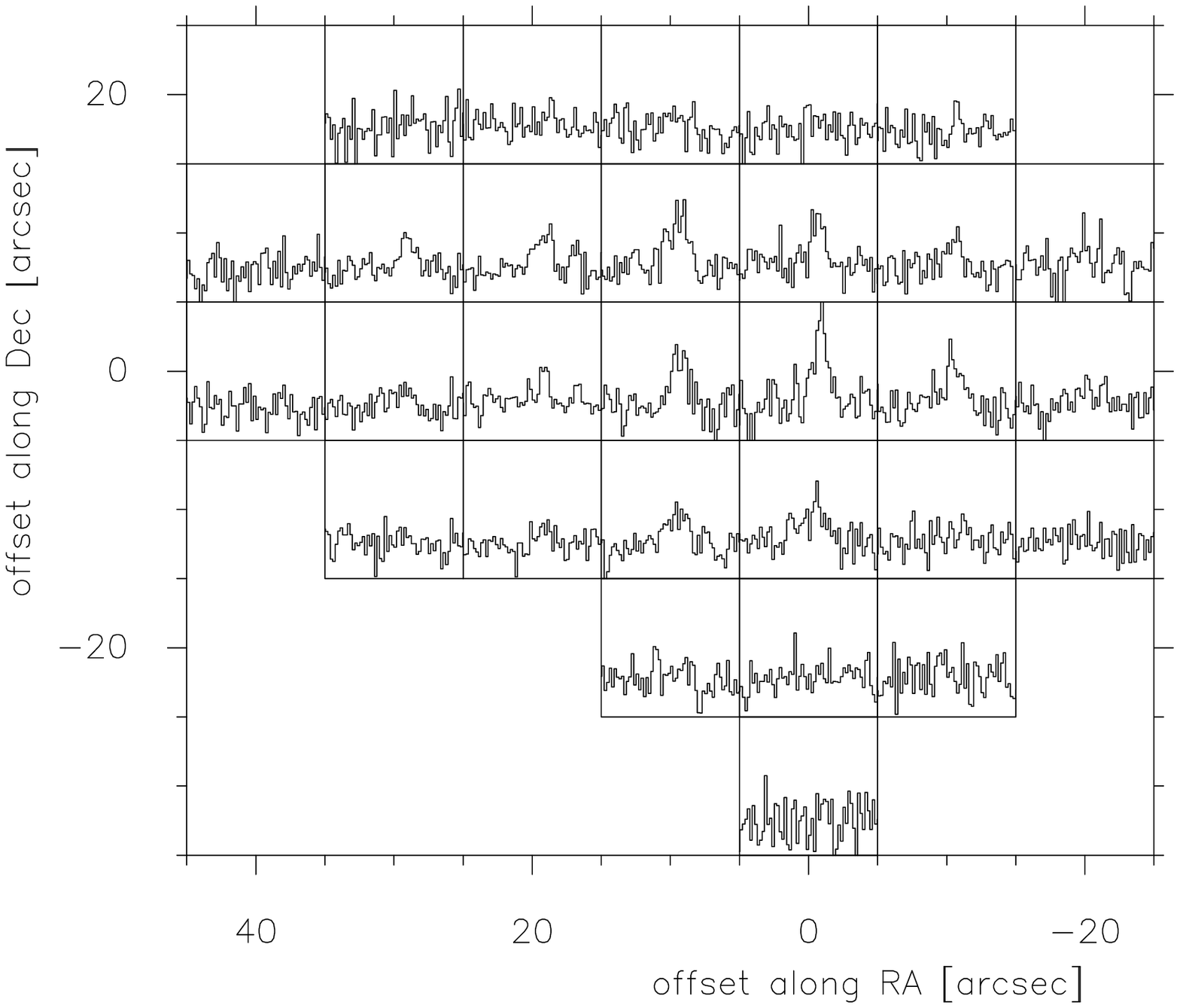}}} 
\caption{CO(1--0) emission in SQ~B, with a velocity resolution
of 5.3\,\kms. The x-scale in the individual spectra 
is velocity and ranges from 6450 to 6750\,\kms.
The y-scale is in $T_{\rm mb}$ and ranges from -13\,mK to 38\,mK.
The noise is on average 4 mK.
}
\label{map_co_one_b}
\end{figure}

\begin{figure}
\resizebox{\hsize}{!}{\rotatebox{270}{\includegraphics{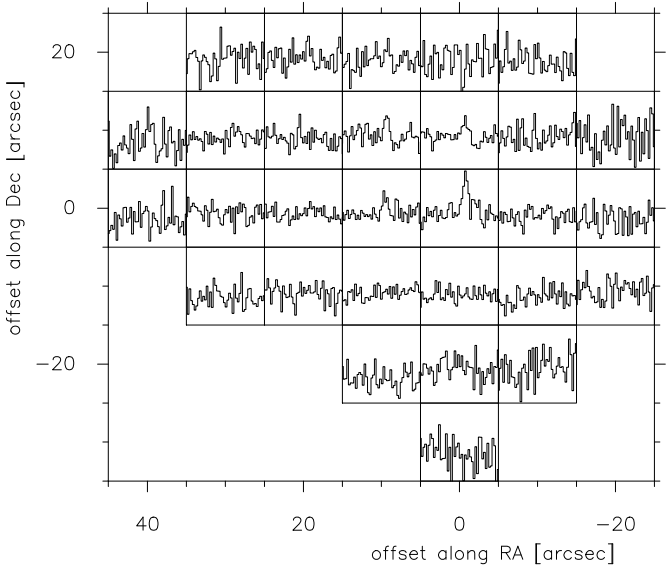}}} 
\caption{CO(2--1) emission in SQ~B, with a velocity resolution
of 5.0\,\kms. The x-scale in the individual spectra 
is velocity and ranges from 6450 to 6750\,\kms.
The y-scale is in $T_{\rm mb}$ and ranges from -34\,mK to 50\,mK.
The noise is on average 5--6 mK.
}
\label{map_co_two_b}
\end{figure}

In Fig.~\ref{map_co_one_a} and Fig.~\ref{map_co_one_b} we present a mosaic
of the CO(1--0) spectra observed around SQ~A and SQ~B.  The noise in the
spectra is about 6\,mK (SQ~A) and 5\,mK (SQ~B) at a velocity
resolution of 2.66 km s$^{-1}$.  
{  The CO(2--1) emission was also
detected at several positions where the \nobreak{CO(1--0)} line was visible
(see Fig.~\ref{map_co_two_b}).
The quality of the CO(2--1) map of SQ~A is much poorer, so that no
useful information can be derived from the individual spectra.
Therefore we do not show it here. 
}
In Table~\ref{tab_mh2_hi_sfr} the integrated intensities
and the derived molecular gas masses for the two regions are
listed. All molecular gas masses in this paper are calculated using
the galactic conversion factor of  
N(H$_2$)/I$_{CO} = 2\times 10^{20}$ cm $^{-2}$(K km s$^{-1}$)$^{-1}$,
yielding:

\begin{equation}
M_{\rm H_2} [M_\odot] = 75 I_{\rm CO} D^2 \Omega ,
\end{equation}
where $I_{\rm CO}$ is the velocity integrated CO line intensity expressed
in K \kms, $D$ is the distance in Mpc and $\Omega$ is the area covered
by the observations in arcsec$^2$ (i.e. $\Omega=1.13 \Theta^2$ 
for a single pointing with  a Gaussian beam of FWHM $\Theta$).
{  
We direct the reader to \citet{Braine01} for a more detailed derivation of the
molecular gas mass. 
}
When referring to molecular data from other authors, we scale their
values to this conversion factor throughout this paper.

{ 
The CO emission in SQ~A is very extended. CO lines at a central
velocity of about 6000 \kms \
are visible in individual spectra over an area of about 40$'' \times$
50$''$ (Fig.~\ref{map_co_one_a}).
The total molecular gas mass at this velocity,
derived  by integrating over the entire area observed, 
is $2.2  \times 10^9$\,\msun.
If we were to limit the integration to the
positions where the CO(1-0) line is clearly visible,
we would obtain an integrated intensity of
$1.2 \pm 0.06$ K km s$^{-1}$ and a total molecular gas mass
of $1.3 \times 10^9$\,\msun. The error in these
masses is about  10\%, partly due to the error in the integrated
intensity (5\%) and partly due to the error in the 
total area which we estimate to be about 5 -- 10\%.
Thus, by summing over those spectra with clearly detected signal we
apparently miss a faint, extended component which
accounts for $40 \pm 6$\%   of the total mass present.
%
}
In the integrated CO spectrum a weaker
component at 6700 \kms \, is visible as well (Fig. \ref{aver_one_a}).
This component is not apparent in the individual spectra, and therefore
we do not include this velocity range in Fig. \ref{map_co_one_a}. 
Its total molecular mass, integrated over the whole area, is
 $8.6 \times 10^8$\,\msun. 

In SQ~B the molecular gas has been detected over a region
of about 60$'' \times$ 40$''$  (Fig.~\ref{map_co_one_b}). It is 
associated with the optically visible star forming region 
-- the maximum of the
CO emission coincides with it -- but  is more
extended. The CO emission includes the region
at the very tip of the optical tidal tail 
where weaker H$\alpha$ and a stellar condensation with blue 
colour  \citep[$B-V=0.2$;][]{Sulentic01} has been found. 
The total molecular gas mass in this area ($7.0 \times 10^8$\,\msun,
see Tab.~\ref{tab_mh2_hi_sfr})
is more than twice as much
as that found by  \citet{Braine01} who observed
only the central position.

We did not detect CO at the majority of the positions of the 
TDG candidates observed by \citet{Mendes01} 
south and north of NGC~7318a/b.  We got a marginal detection 
of $I_{\rm CO} = 0.5$ K\, \kms \,  when averaging
the positions SQ~C16/17 and  SQ~C18/19  which we
used as an upper limit. 
No CO was detected in the northern part of the eastern HI tail
(SQ~B-North). This position was chosen because it coincides
with a local peak of the HI emission and shows \halpha \ emission
\citep{Sulentic01}.
The rms noise of the positions with non-detections are
3--4\,mK for CO(1-0) and 4--6\, mK for CO(2-1), at a velocity resolution
of 5 \kms.


\begin{table*}
\begin{minipage}{18cm}
\caption{Molecular, atomic and ionized gas in the different regions observed}
\begin{tabular}{ccccccccc}
\hline
Region & $I_{\rm CO}$ & Size(\mhtwo) & \mhtwo$^{(1)}$  & \mhi$^{(2)}$ & \mhtwo/\mhi  & 
L(\halpha) & SFR$^{(3)}$ &  \mhtwo/SFR \\
 & [K \kms] & [arcsec$^2$] &  [10$^8$ \msun] & [10$^8$\msun] &   & 
[10$^{40}$erg s$^{-1}$]  &  [\msun/yr]  & 10$^{8}$ yr \\
\hline
SQ~A(6000)$^{(4)}$& 0.84$\pm$0.04 & 60$\times$80 & 22 & 16 & 1.4 & 8.8/26.4$^{(8)}$ & 1.16/3.48  & 19/6 \\
SQ~A(6700)$^{(5)}$& 0.33$\pm$0.05 & 60$\times$80 & 8.6 & 9  & 0.95 & 2.2/6.6$^{(8)}$  & 0.29/0.86  & 30/10 \\
SQ~B(tot)$^{(6)}$ & 0.54 $\pm$ 0.02 & 60$\times$40 & 7.0  & 14 & 0.5 & 1.2$^{(8)}$ & 0.16  & 44 \\
SQ~B(cent)$^{(7)}$ & 1.1$\pm$0.1   & 21$\times$21 & 2.9$^{(10)}$  & 5  & 0.6 & 1.2$^{(8)}$ & 0.16    & 18\\
SQ~B-North  & $<$0.09  & 21$\times$21 & $<$0.24 & 2.5 & $<$0.10 & 0.5$^{(8)}$  & 0.06 &  $<$4.0\\
SQ~C8        & $<$0.09 & 21$\times$21 & $<$0.24  & 0.7 & $<$0.34 & 0.8$^{(9)}$  & 0.10 &  $<$2.4\\
SQ~C16/17-18/19& $<$0.5& 31$\times$21 &$<$1.76  & 5.5 & $<$0.33 & 8.8$^{(9)}$ & 1.14 &  $<$1.6 \\
SQ~C20       & $<$0.07 & 21$\times$21 & $<$0.19  & 0.3 & $<$0.63 & 0.9$^{(9)}$  & 0.12 &  $<$1.6\\ 
SQ~C21     & $<$0.07  & 21$\times$21 & $<$0.19  & 1.1 & $<$0.17 & 1.5$^{(9)}$  & 0.20 &  $<$1.0\\
SQ~C22       & $<$0.05 & 21$\times$21 & $<$0.14 & 1.6 & $<$0.09 & 1.3$^{(9)}$  & 0.17 &  $<$0.8\\
SQ~C23        & $<$0.07 & 21$\times$21 & $<$0.19 & 1.6 & $<$0.12 & 0.4$^{(9)}$  & 0.05 & $<$3.8\\

\hline

\end{tabular}

$^{(1)}$ Calculated using the Galactic conversion factor 
of N(H$_2$)/I$_{CO} = 2\times 10^{20}$ cm $^{-2}$(K km s$^{-1}$)$^{-1}$.

The upper limits are calculated as
$\sigma (\triangle V /\delta V)^{1/2}$, where
$\sigma$ is the rms noise, $\delta V$  the velocity resolution
(here 5.3 \kms) and $\triangle V$  the line width, adopted to 
be equal to the  HI line widths in this region of about 100 \kms \citep{Williams02}.

$^{(2)}$ From HI data of \citet{Williams02}.

$^{(3)}$ Calculated from the \halpha \, luminosity according to
$SFR = 5 \times 10^{-8} (L_{\rm H\alpha}/L_\odot) M_\odot \rm{yr}^{-1}$
\citep{Hunter86}.

$^{(4)}$ Only the gas components
at 6000 \kms.

$^{(5)}$ Only the gas component
at  6700 \kms. The spatial extent cannot be
quantified because the signal is too weak to show up in individual
spectra. The value
listed here refers to the area over which the line was
integrated.

$^{(6)}$ Values for the CO and HI for the entire area with CO emission. 

$^{(7)}$ Values for the CO and HI for the
area within the central beam, coinciding with the 
\halpha \, emission.

$^{(8)}$ From \halpha \, data of \citet{Xu99}.
The first value refers to the luminosity in the starburst region,
the second value to the total luminosity in the region where CO
was mapped.

$^{(9)}$ From \citet{Mendes01}, scaled down by a factor of 2.8 in order
to achieve agreement with the data by \citet{Xu99}.

{ 
$^{(10)}$ The difference of \mhtwo \ with respect to the 
value reported in \citet{Braine01}
is due to the different adopted distances  and the fact that \citet{Braine01}
include the helium mass in their estimate of the mass of molecular clouds. 
}

\label{tab_mh2_hi_sfr}
\end{minipage}
\end{table*}


\subsection{Kinematics and excitation}

\begin{figure}
\resizebox{\hsize}{!}{\rotatebox{270}{\includegraphics{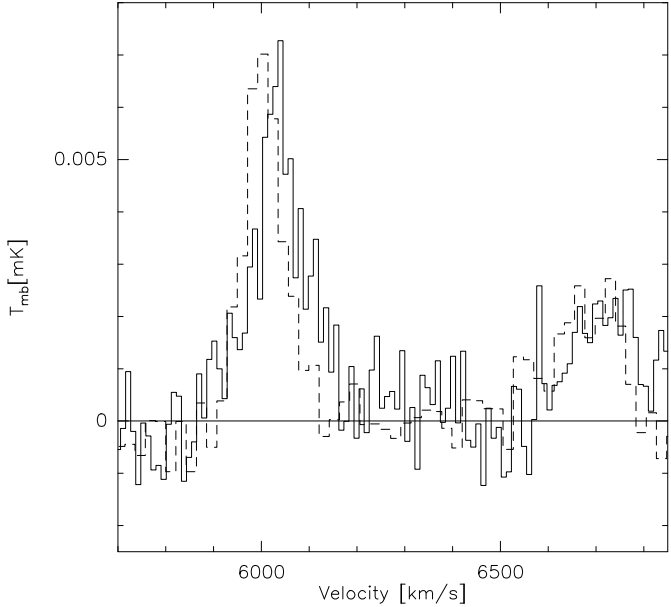}}} 
\caption{CO(1--0) spectrum (full line) in SQ~A, 
{  with a velocity
resolution of 10.6 \kms,} averaged over 
the total observed area. The dashed line represents the HI emission
{  with a velocity resolution of 21.4 \kms \ } in arbitrary units
from Williams et al. (2002), averaged over the same
area.
}
\label{aver_one_a}
\end{figure}

\begin{figure}
\resizebox{\hsize}{!}{\rotatebox{270}{\includegraphics{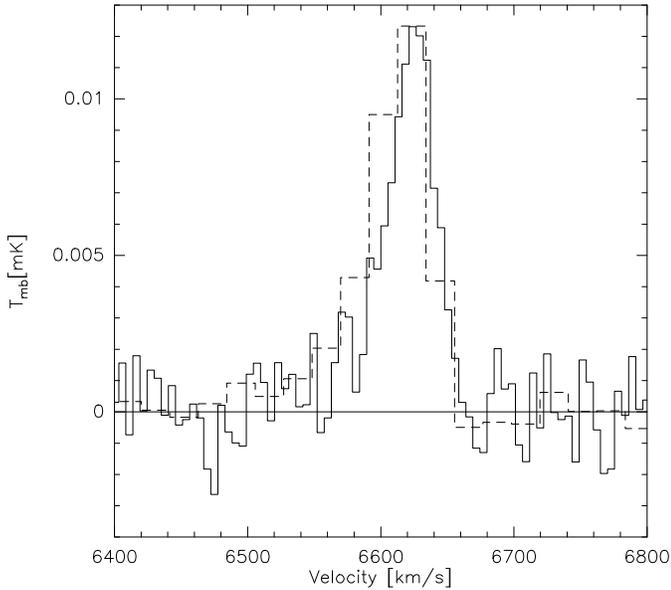}}} 
\caption{CO(1--0) spectrum (full line) in SQ~B,  {  
averaged over the positions
with offsets between 30\arcsec \ and -10\arcsec \ in RA and
-10\arcsec \ and 10\arcsec \ in Dec. The velocity resolution is
5.3 \kms.} 
The dashed line represents the HI emission
{  with a velocity resolution of 21.4 \kms \ } in arbitrary units 
from Williams et al. (2002), averaged over the same
area.
}
\label{aver_one_b}
\end{figure}

In Fig.~\ref{aver_one_a} and Fig.~\ref{aver_one_b}
we show  the average CO(1--0)
spectra in both regions, overlaid over the HI emission 
\citep{Williams02}, averaged over the same region.
The line shapes in both regions are different:
In SQ~A the CO line is very broad,
with a FWHM of $60-80$\,\kms \, and line wings extending out to
more than 200\,\kms, showing that in  this region
large velocity gradients are present, whereas the line width
in SQ~B is much narrower, with a FWHM of $30-40$\,\kms.

The HI gas of SQ~A has two distinct velocity
components, one at 6000 and another at 6700\,\kms. 
The CO and HI lines generally agree well,
with the exception of a slight velocity offset 
between the low-velocity component of the HI line, 
centered at 6000 \kms, and the CO line, centered at 6030--6040 \kms.
Nevertheless, the line width, shapes and
the relative intensity of the components at 6000 and at 6700 \kms \ are 
the same for the HI and CO spectra.
\citet{Smith01} find, using the NRAO 12m telescope, that
both velocity components have similar strengths. The reason for
why they underestimate the strength of the line at 6000 \kms
is most likely because they cover a smaller area. 
%
%
This can be seen when comparing the 
molecular gas masses that they derive for their components. 
Whereas their value for the component at 6700 \kms, 
\mhtwo$= 7 \times 10^8$ \msun, agrees well with ours,
their value for the component at 6000 \kms, 
\mhtwo$= 7.9 \times 10^8$ \msun, is much lower.
This can be understood if the 6700 \kms \ component
is more centrally concentrated, 
{  as is the case for the HI emission \citep{Williams02}},
and as
a result is covered by their beam,
but the 6000 \kms \ component is not.
From their Fig. 1g it is indeed
obvious that their beam is smaller than the area mapped
by us.

The CO emission in SQ~B has a central velocity of 6625\,\kms.
There is very good agreement between the CO
and HI lines \citep[from][]{Williams02} with respect to
the line width, central positions and 
special features, such as the 
blue wing.

\begin{figure}
\resizebox{\hsize}{!}{\rotatebox{270}
{\includegraphics{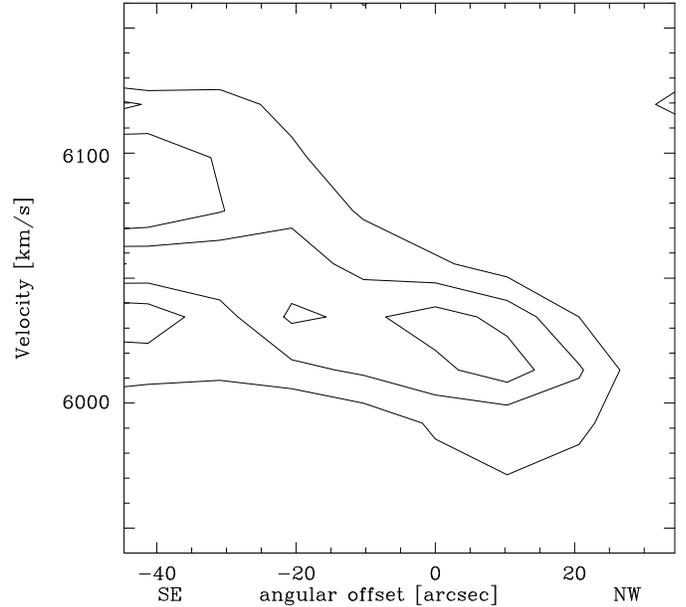}}} 
\caption{Position-velocity diagram of the CO(1--0) emission in 
SQ~A at a position angle of $-40$ degrees. The contours are 
at 4.5, 7 and 9.5 mK ($T_{\rm mb}$).
}
\label{sqa_pv_pa325}
\end{figure}


To better understand the dynamical behavior of the CO emitting gas we
created several position-velocity diagrams using our data.  Our
analysis of SQ~B only gave marginal 
indications for a weak velocity gradient of about 
$10 - 20$ \kms \, along the east-west direction.
In SQ~A we did
find evidence for a slope in velocity along a position angle of
$-40$ degrees  for the velocity
range around 6000 \kms \ (Fig.~\ref{sqa_pv_pa325}). 
Moreover, for a cut perpendicular
to it, no slope was seen. 
This is very similar to what we found for the
atomic gas using the HI data of \citet{Williams02}. 
A similar trend has also been seen in the
\halpha \  data of \citet{Mendes01} for their regions 2,3,4 and 5. 
The position angle of the \halpha \ velocity
is the same as the one we find
in CO and HI, but their  observed velocity range is slightly 
smaller, between 5938\,\kms \
and 6010\,\kms, whereas the CO data shows a gradient  from 
about 5970\,\kms \ to
6130\,\kms. This is probably due to the smaller extent of only about
30\arcsec \ of the \halpha \ observations. 
Normalized to the same spatial extent
the \halpha \ gradient is comparable to the gradient in CO within the
uncertainties.
%

\begin{figure}
\resizebox{\hsize}{!}{\rotatebox{270}{\includegraphics{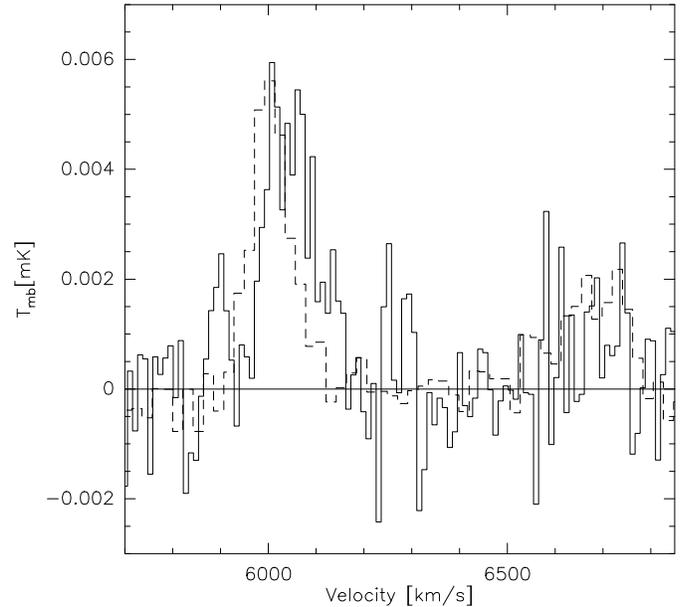}}} 
\caption{CO(2--1) emission (full line) in SQ~A,  averaged over
the total observed area. {  The 
velocity resolution is 10.6 \kms.} The dashed line represents the HI emission
{  with a velocity resolution of 21.4 \kms \ } in arbitrary units
from Williams et al. (2002), averaged over the same
area.
}
\label{a_two_aver}
\end{figure}

\begin{figure}
\resizebox{\hsize}{!}{\rotatebox{270}{\includegraphics{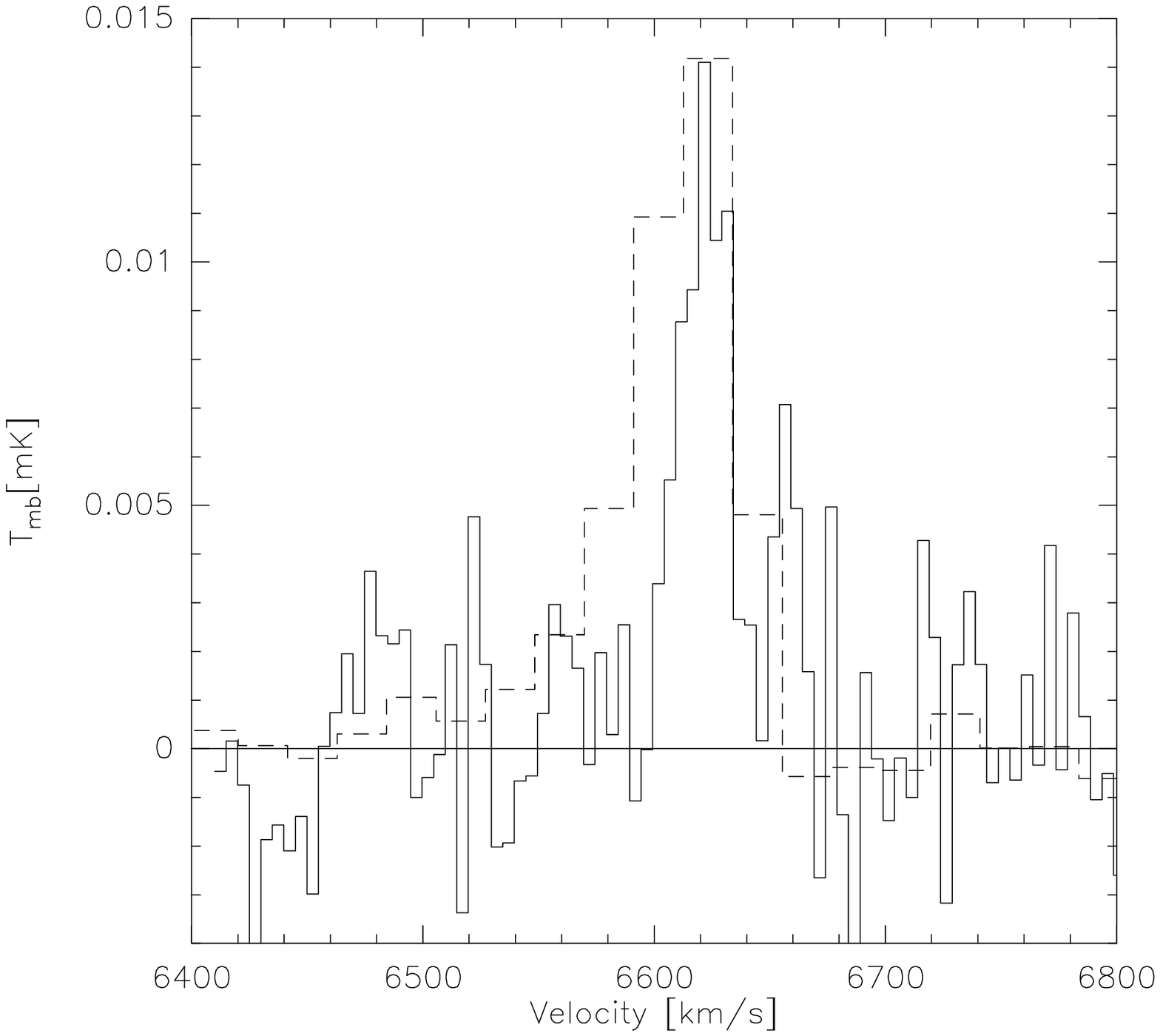}}} 
\caption{CO(2--1) emission (full line) in SQ~B, 
{  averaged over 
the same area as in Fig. \ref{aver_one_b} and with a 
velocity resolution is 5.0 \kms.} 
The dashed line represents the HI emission
{  with a velocity resolution of 21.4 \kms \ } in arbitrary units
from Williams et al. (2002), averaged over the same
area.
}
\label{b_two_aver}
\end{figure}

In Fig.~\ref{a_two_aver} and \ref{b_two_aver} we show the 
\nobreak{CO(2--1)} lines, {  averaged over the same area as the 
corresponding \nobreak{CO(1--0) lines.} 
}
%
{  
In order to compute the CO(2--1)/CO(1--0) 
integrated intensity ratio in SQ~A and SQ~B, 
the CO(2--1) spectra have been convolved to the spatial resolution of 
the CO(1--0) beam. To minimize the error of this ratio, we have
included only spectra at positions with a  S/N $>3 \sigma$
of the CO(1-0) line. 
In SQ~A, the 
CO(2--1)/CO(1--0) ratio is $0.69 \pm 0.16$ and
in SQ~B $0.56 \pm 0.13$. 
Within the errors, there is no significant difference in the
line ratios. They are slightly lower than the average found for
a mixed sample of nearby galaxies 
\citep[CO(2--1)/CO(1--0)$ = 0.89 \pm 0.06$,][]{Braine92}.
Better CO(2--1) data are necessary in order to judge whether this
difference is significant.
}

\subsection{Comparison of the CO and HI distribution}

\begin{figure}
\resizebox{\hsize}{!}{\rotatebox{270}
{\includegraphics{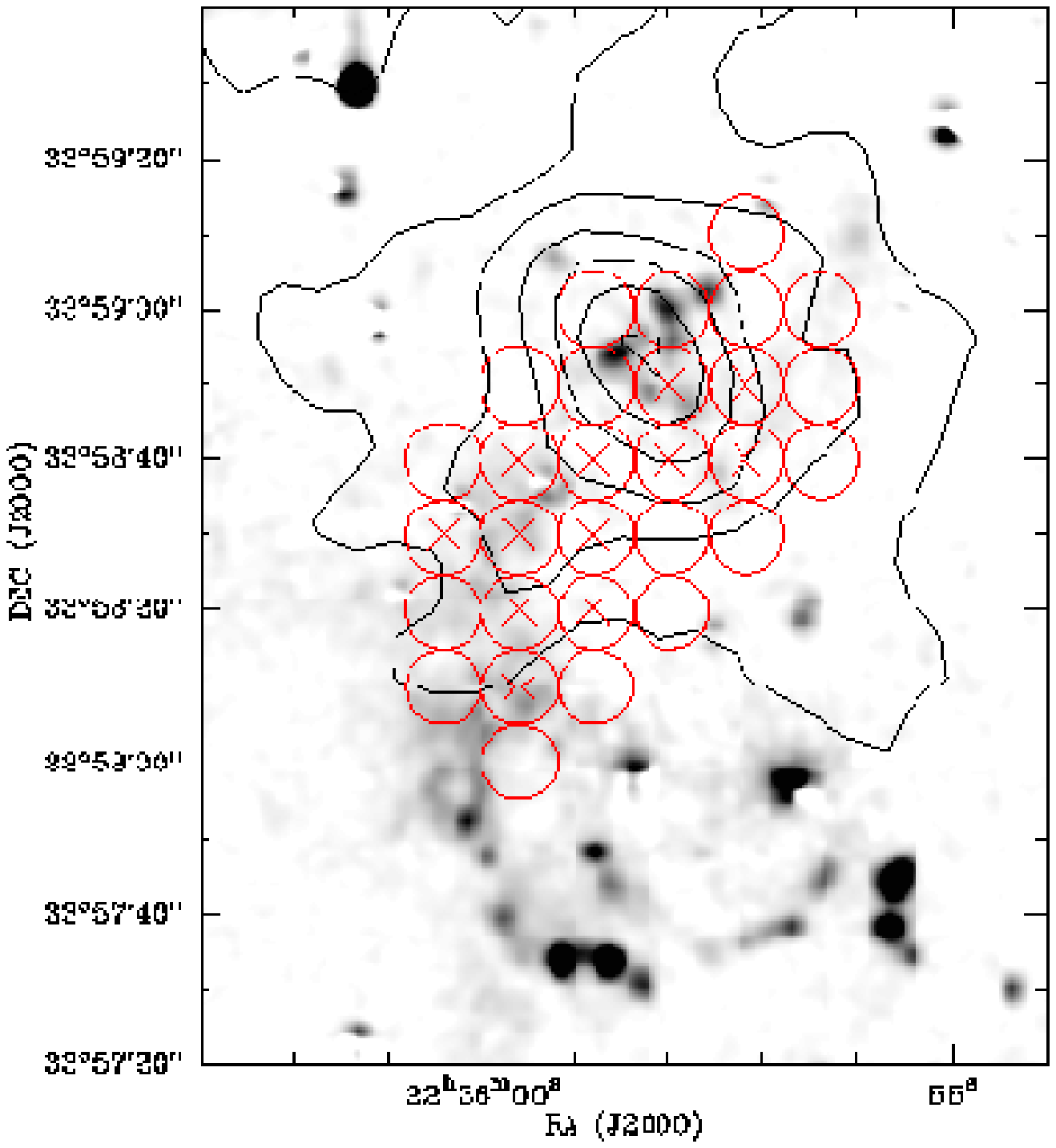}}}
\caption{A grey scale image of the \halpha \, emission in SQ~A
obtained with a filter centered at 6667 \AA \, and a FWHM of
66 \AA \, \citep[from][]{Xu99}, covering the \halpha \, emission around
6000 \kms,  
overlaid with an HI contour map \citep[][levels at $1.2, 2.4, 3.6, 
4.8, 6.0, 7.2, 8.4 \times 10^{20}$ atoms]{Williams02}
showing the velocity integrated column density 
of the line at 6000 \kms.
The circles show the location of our pointings and the crosses indicate the
positions where CO emission was detected above a level of $ 3 \sigma$.
}
\label{sqa_hi_ha_co}
\end{figure}

\begin{figure}
\resizebox{\hsize}{!}{\rotatebox{270}
{\includegraphics{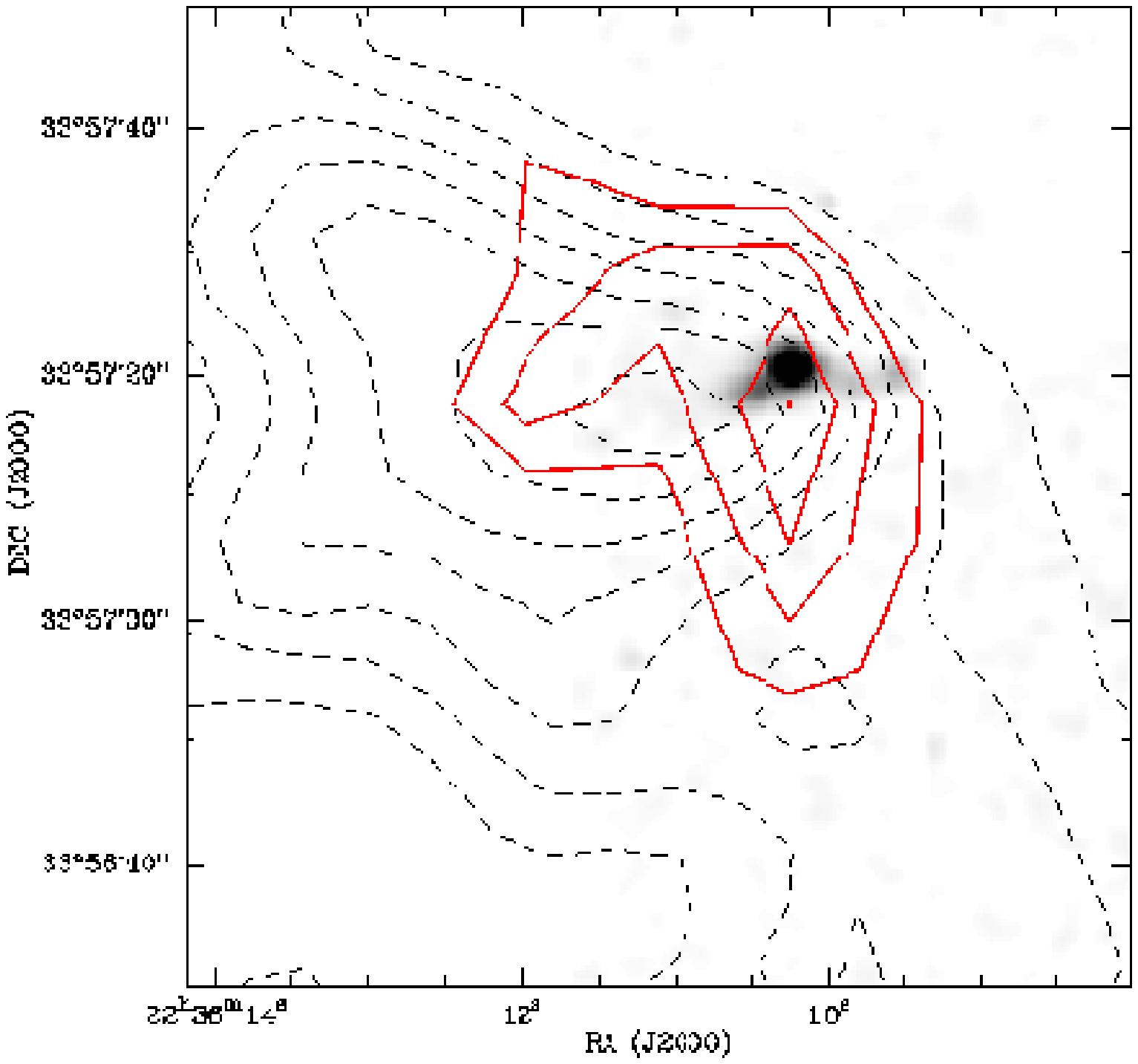}}} 
\caption{A grey scale image of the \halpha \ emission in SQ~B
obtained with a filter centered at 6737 \AA \, and a FWHM of
76 \AA \, \citep[from][]{Xu99}, covering the \halpha \, emission around
6600 \kms, 
overlaid with an HI contour map \citep[][dashed contours, levels 
at $1.2, 2.4, 3.6, 4.8, 6.0, 7.2, 8.4 \times 10^{20}$ atoms \cm]{Williams02}
showing the velocity integrated column density 
of the line at 6600 \kms, and
thick (red) contours representing $I_{\rm CO}$, 
the velocity integrated CO emission,
(contour
levels at 0.4, 0.65, 0.9 and  1.15 K km s$^{-1}$).
}
\label{sqb_hi_ha_co}
\end{figure}

In spite of the practically identical kinematics, the atomic and molecular gas
distributions show differences both in SQ~A and in SQ~B.
In SQ~A, the CO lines are wide and 
have low maxima. As a result, it is difficult to display
the velocity integrated intensities of the CO emission
because of their large uncertainties.
We therefore show in Fig.~\ref{sqa_hi_ha_co} a comparison of
the HI emission around 6000 \, \kms \, with the observed positions in CO.
The positions where we clearly ($>3 \sigma$) detected a
line are indicated by a cross and the observed positions as circles.
Note that, as mentioned earlier, also at the positions without a clear
CO detection some faint emission must be present.
From Fig.~\ref{sqa_hi_ha_co} it is clear that 
the overall distribution of the CO 
emission is different from the HI
emission.  Even though some CO marginally
coincides with the HI peak, most of it is located to the south-east
along a faint HI ridge.

In Fig. \ref{sqb_hi_ha_co} we display the contours
of the integrated CO intensity
of SQ~B overlayed onto contours of an HI 
integrated surface brightness map.
The maxima of the CO and HI emission are close, but do
not exactly coincide.
Instead, the CO peaks slightly (10 --  15 \arcsec) west from the HI ridge,
towards a steep HI gradient, suggestive of compression in the gas
which implies that this over-density in HI may have provoked the
formation of molecular gas.  The molecular-to-atomic gas ratio is
highest at the peak of the CO emission and west of it. This means,
under the assumption that the molecular gas is being formed in situ,
{  as has been suggested for TDGs \citep{Braine01}},
that this compressed region is the place of the most efficient
transformation of HI into H$_2$.

The HI mass in SQ~A, integrated over the area mapped in CO,
is $1.6 \times 10^9$ \msun ($9 \times 10^8$ \msun) for the
velocity component at 6000 \kms \, (6700 \kms) 
(see Tab.~\ref{tab_mh2_hi_sfr}). 
{  This means that the molecular gas mass
is comparable to the atomic gas mass with a ratio of \mhtwo/\mhi =
1.4 (1.0)}. 
The HI mass of the entire area around 
SQ~A is $3.1 \times 10^9$ \msun, which is of the 
same order as the total mass of the gas in molecular form.
The atomic gas mass in SQ~B, integrated over 
the region where CO was detected, is $1.4
\times 10^9$\,\msun.
This yields a molecular-to-atomic gas mass ratio
of 0.5, higher than the values of less than about 0.2 found
for other TDGs \citep{Braine01}.

\subsection{Comparison of the CO and H$\alpha$ distribution}       

The \halpha \ emission in SQ~A 
consists, {  as do the HI and CO data}, 
of two velocity components, 
centered around 6000 \kms \ and 6700 \kms, respectively
\citep{Sulentic01,Iglesias01,Xu99}. The distribution 
of both velocity components
is similar in the regions that our CO observations covered.
Using the 
data of \citet{Xu99}, taken with narrowband filters,
we present in Fig.~\ref{sqa_hi_ha_co} 
the spatial distribution of the \halpha \, emission at  6000\,\kms \,
and our CO detections. We observe that the CO gas
follows closely the H$\alpha$ emission in the north-south direction,
but surprisingly little CO is seen close to 
the brightest H$\alpha$ peak in the starburst region at SQ~A.

Similarly, we present in Fig. \ref{sqb_hi_ha_co} the H$\alpha$ image of
SQ~B overlayed with the contours of the molecular gas.  
The H$\alpha$ emission is situated 5\arcsec \  north of the maximum of the CO
emission and approximately 15\arcsec \ north-west
from the maximum of the HI emission. 
We conclude that the H$\alpha$ \ and CO emission 
coincide within
the limited spatial resolution
(21\arcsec) of our data.
Also, the velocities
agree, with the velocity
of the CO line being 6625 \kms \ and that of 
the \halpha \, emission 6617 \kms \ \citep{Sulentic01}.
Altogether, this is pointing to a causal
connection between molecular gas and recent star formation. 

In Tab. \ref{tab_mh2_hi_sfr} we  list the molecular gas mass
divided by the SFR (derived from the \halpha \ luminosity)
which gives the gas consumption time 
(i.e. the inverse of the star formation efficiency). 
The values we derive for SQ~A and SQ~B are typical to
those of  spiral galaxies, which have been estimated
to lie between 4 and 40 $\times 10^{8}$ yr \citep{Kennicutt98}.

\subsection{Molecular gas fraction and star formation 
in the non-detections}

We observed various positions (see Fig.~\ref{opt_map}) south
of NGC~7318a/b and north of SQ~B in CO without getting a detection.
The results are summarized in Tab. \ref{tab_mh2_hi_sfr}.

The position north of SQ~B (SQ~B-North) coincides with the
north tip of the HI cloud and was chosen because 
\citet{Sulentic01} detected \halpha \ emission at the same velocity as the
HI and suggested that an  extragalactic HII region may be 
forming. \citet{Gallagher01} identified two starclusters (30 and 31)
at the same position.  The clusters have
blue colours and are therefore young, less than
10 -- 500 Myr, and must have formed in situ.
At this position we derived an upper limit of
\mhtwo/\mhi $\le 0.08$.  This value is much lower
than at the other places in SQ where we detected molecular gas,
but it is not unusual compared to typical values for
TDGs \citep{Braine01}.
The upper limit of \mhtwo/SFR$<4 \times 10^8$ yr  is at the low end of the 
typical range for  spiral galaxies. 

Around  NGC~7318a/b we observed several of the positions 
where \citet{Mendes01} have found \halpha \, emission (their positions
8, 16/17, 18/19, 20, 21, 22 and 23; {  note that positions 16 and 17
as well as 18 and 19
fall  within the IRAM beam}). 
The authors suggest that some of them (8, 20, 21, 22 and 23)
are TDG candidates, mainly based on 
the detection of a velocity gradient  in the \halpha \ data
which  they interpret as rotation.
We observed in addition to these 
TDG candidates positions 16/17 and 18/19 because of their 
relatively high HI column density.
The upper limits of the molecular gas fraction, \mhtwo/\mhi, 
are low for some objects, but not unusual compared
to spiral galaxies and TDGs. 
The upper limits for the gas consumption time,
\mhtwo/SFR, on the other hand,
are, for most objects, much lower than the values found for TDGs
\citep[see][]{Braine01} or spiral galaxies.
This is particularly surprising, as the large size of the CO(1-0)
beam of 21\arcsec , corresponding to 8.6 kpc, should comprise
the molecular gas associated with the star formation even if
the \halpha \, and CO emission do not coincide exactly,
because, e.g. the ongoing star formation has destroyed the
molecular gas locally.
A low metallicity could explain the non-detections
but seems unlikely as we did detect CO at
other places in SQ. 
We checked for the presence of star clusters from
\citet{Gallagher01} at the observed
positions 
{  in order to find out whether star clusters have been formed
since the beginning of the collision with NGC~7318b 
of less than $10^8$ yr ago \citep{Sulentic01}.}
Region C23 coincides with an old star-cluster
(number 117), with an age
of several Gyr, which was therefore born before the start of the latest
interaction. 
Regions C8, C16/17 and C18/19 coincide with 
star clusters 133, 134, 135, 137, 122, 
128, and 130, some of which are younger than 10 Myr.
At the other observed positions no star clusters were
found within a radius of 10\arcsec. 
{  Thus, young star clusters formed  after
the beginning of the collision have  only been found close to  
the parent galaxy NGC~7318a and not in the TDG candidates 
C20 -- C23.}

\subsection{Comparison of the CO distribution to other tracers}

\begin{figure}
\resizebox{\hsize}{!}{\rotatebox{270}
{\includegraphics{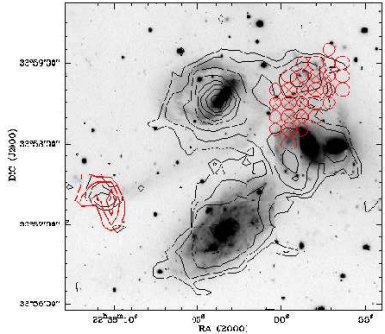}}} 
\caption{A grey scale optical image with a contour overlay of the 15 $\mu$m
emission by ISOCAM \citep{Xu99}. The thick (red) contours on SQ~B and the 
circles on SQ~A show our CO observations, as described in 
Fig. \ref{sqa_hi_ha_co} and \ref{sqb_hi_ha_co}.
}
\label{sqab_iso_co}
\end{figure}

\begin{figure}
\resizebox{\hsize}{!}{\rotatebox{270}
{\includegraphics{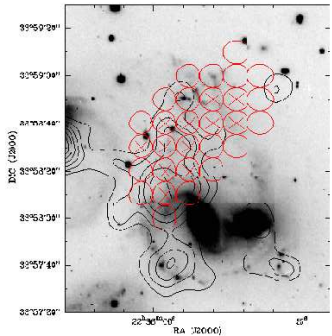}}} 
\caption{A grey scale optical image overlaid with contours of the
X--ray HRI emission from ROSAT \citep[see][]{Pietsch97}, 
smoothed to a resolution of 10\arcsec. 
The circles on SQ~A show our CO observations, as described in 
Fig. \ref{sqa_hi_ha_co}. At SQ~B, no X--ray emission was 
found.
}
\label{sqab_xray_co}
\end{figure}

\begin{figure}
\resizebox{\hsize}{!}{\rotatebox{270}
{\includegraphics{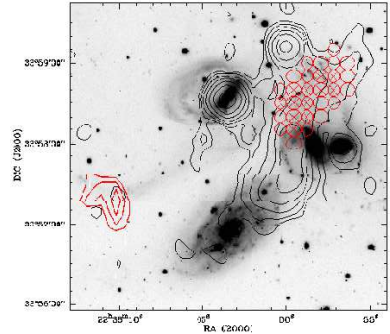}}} 
\caption{A grey scale optical image overlaid with contours of the
20cm radio continuum emission \citep{Williams02}. 
The thick (red) contours on SQ~B and the 
circles on SQ~A show our CO observations, as described in 
Fig.  \ref{sqa_hi_ha_co} and \ref{sqb_hi_ha_co}.
}
\label{sqab_cont_co}
\end{figure}

In Fig. \ref{sqab_iso_co}, \ref{sqab_xray_co}  and \ref{sqab_cont_co}
we compare the CO distribution to that of the 15 $\mu$m emission
observed by ISOCAM \citep{Xu99}, the X--ray emission from the
ROSAT archive, and the 20cm radio continuum emission
\citep{Williams02},  respectively.

The overall distribution of the dust traced by the ISOCAM
15  $\mu$m image is similar to the CO emission in both SQ~A
and SQ~B (Fig. \ref{sqab_iso_co}). In SQ~A the
15  $\mu$m emission coincides with the maximum in the
\halpha \ and indicates the position of the intergalactic
starburst. The CO is slightly offset towards the
south and west of the 15  $\mu$m peak. 
The CO emission 
coincides with faint 15  $\mu$m emission at all our
observed positions, showing that the 
presence of molecular gas is accompanied by the presence of 
dust.
In SQ~B the 15  $\mu$m
peak is marginally offset (5-10\arcsec) from the
CO maximum and is located towards the peak of the HI surface
density.

In the intergalactic medium around SQ~A, radio continuum 
\citep{Williams02} and X--ray emission \citep{Pietsch97}
are present. As previously noted \citep[e.g.][]{Sulentic01},
there is  good agreement between the radio continuum, the X--ray and
\halpha \, emission in the elongated north-south feature 
north of NGC~7318a/b.
All along this ridge we find CO emission 
(Fig. \ref{sqab_xray_co} and
\ref{sqab_cont_co}). 
{  
CO is also present in a region without  X--ray emission
towards the north
which coincides with the edge of the ISOCAM peak.
}
In SQ~B no X--ray emission is detected, but weak ($2\sigma$) 
radio continuum
emission can be seen, coinciding with the peak of the CO emission.

The radio continuum emission in both regions is 
mostly non-thermal (i.e.
synchrotron emission) 
as can be deduced by calculating the expected thermal radio
emission from the \halpha \ emission.
The thermal radio emission derived   
from the \halpha \, flux yields 
\citep[following eq. (3) and (4) in][]{Condon92}, a few percent of the 
radio continuum flux in SQ~A and about 10\% in SQ~B.
This thermal fraction at 1.49 GHz
is typical  for star-forming galaxies \citep{Condon92}.

\section{Discussion}

\subsection{Molecular gas, shocks and star formation in SQ~A}

The enormous amount of molecular gas that we have found
and its large extent are 
surprising. 
In the following we will  discuss how this molecular
gas fits into the current picture of 
the IGM in SQ~A.

The IGM around SQ~A shows  two velocity components.
\citet{Sulentic01} suggest that the gas at 6700 \kms \
was stripped 
during the earlier visit  of  NGC~7320c 
and part of it has become shock-ionized due to the present interaction
with NGC~7318b.
In this view, the \halpha \, emission at 6700 \kms \
traces the shock front.
Strong support for this shock hypothesis is the fact that
X--ray (tracing hot gas) and radio continuum emission (tracing cosmic ray
electrons produced in shocks) are coincident with the H$\alpha$
emission.  
The gas at velocities around 6000 \kms \ originates from
NGC~7318b and is being removed from this galaxy in the ongoing
collision.
The  \halpha \ emissions at 6000 and 6700 \kms \ 
overlap to a large extent in the region where we observed
CO. 
{  In HI, the two velocity components have their maxima at the same
position, but the component at 6700 \kms \ is spatially much less 
extended than that at 6000 \kms \ \citep{Williams02}.
} 
\citet{Sulentic01}
suggest that the apparent spatial 
coincidence is a projection and that NGC~7318b, and thus the 
shock caused by it, are  moving along the line of sight.

The presence of molecular gas indicates a rather cool environment. 
It is therefore surprising to find large amounts and 
extended CO emission at positions where 
shocks are present and are heating and ionizing
the gas.  J-shocks with velocities above about 50 \kms \ are able
to effectively destroy molecular gas \citep[][]{Hollenbach80,
Hollenbach89}. Shocks with these velocities are plausibly
present given the large velocity difference (600 -- 1000 \kms) between
the intruder NGC~7318b and the rest of the SQ galaxies.

This strongly suggests that shocks do not affect the gas at 6000 \kms,
the velocity at which we found the major part of the 
CO emission. 
This means that the X--ray and radio continuum emission 
tracing shocks are only associated with the gas at 6700 \kms.
In addition to the presence of extended  molecular gas, 
support for this hypothesis comes from the atomic gas.
{  
Similarly to the CO,
the HI component at 6000 \kms \ is 
very extended,  indicating that shock heating
is not affecting this area so that cool gas can exist.
Contrary to this, the 
HI component at 6700 \kms \ is spatially much more concentrated.
}
Furthermore, the spatial coincidence between the 
15 $\mu$m and  and CO emission
support the absence of shocks because dust is very easily
destroyed by them \citep[e.g.][]{Borkowski95, Jones96}.

{  
The molecular gas mass in the starburst region\footnote{ 
I.e. 
adding the 4 pointings around
the position (-5\arcsec, 5\arcsec), where the maximum of the
\halpha \ emission is found, and summing both
velocity components.
} 
is $7.0 \pm 1.0 \times 10^8$ \msun.
The surprising result of our observations
is that the molecular gas content of the starburst region is
only a small part (about one fourth) of the total molecular gas
content and that the starburst does not mark the 
maximum of the CO emission. 
Our molecular gas mass in the starburst region 
is only slightly higher than the mass estimated from 
interferometric observations by \citet{Gao00}   
($2.4 - 5.3 \times 10^8$ \msun).
The observations of
\citet{Gao00} showed CO emission only at the position
of  the starburst. This
could indicate that the extended emission that we have found is
more smoothly distributed and therefore not detected by  
interferometric observations. 
}

The origin of the molecular gas in SQ~A is unclear. It could have been
stripped together with the atomic gas from NGC~7318b or it
could have formed in situ. In the center of NGC~7318b, 
a molecular gas mass of  $1.2 \times 10^9$ \msun \
has been found \citep{Smith01}. This yields, together with its
blue luminosity of $4.6 \times 10^{10}$ \msun, a ratio of
$L_{\rm B}/M_{\rm H_2} = 38$, considerably 
higher than the 
{  average value of
$L_{\rm B}/M_{\rm H_2} \approx 6$ (with a scatter of a factor 3)
that has been found
}
for isolated galaxies of this luminosity
\citep{Perea97}. Thus, the molecular gas
in SQ~A could have its origin in NGC~7318b.
On the other hand, if the molecular gas has formed
in situ, a large scale process and not gravitational
collapse must be responsible for its formation over a scale of about
25 kpc.
The velocity of this gas, 6000 \kms \ (which is higher than
the central velocity of NGC~7318b of 5765 \kms) could be an
indication that it has already been accelerated by the
collision with the intergalactic gas and the slight offset
between the molecular (6030 \kms) and atomic gas (6000 \kms) 
central velocities  might show that the molecular gas is forming
in the most accelerated parts.

\subsection{Origin of the molecular gas in SQ~B}

{  The molecular gas in SQ~B could have either been
stripped  from the parent galaxy 
together with the atomic gas
or it could have formed in situ. 
%
%
Studying a sample of TDGs,
\citet{Braine01} argue in favor of the in-situ formation.
}
Their main argument  is that
the CO distribution in spiral galaxies is normally 
concentrated towards their central
parts whereas the atomic gas  is much more widely distributed 
and therefore more easily stripped.
In SQ the situation is more complicated because more than two 
galaxies are interacting. However, the 
arguments of \citet{Braine01} seem to be applicable to NGC~7319,
the galaxy thought to be the progenitor of the 
eastern HI cloud and the young optical 
tidal tail. 
All of the HI from this galaxy has been stripped, 
whereas CO is still found within its optical disk
\citep[][]{Yun97,Smith01}. 
Its molecular gas content of $4.8 \times 10^9$ \msun \
\citep{Smith01} and its blue luminosity
\citep[$L_{\rm B}= 5.5 \times 10^{10}$ \lsun;][]{Verdes98}
yield $L_{\rm B}/M_{\rm H_2} = 11$ which is normal compared
to other galaxies (see above in Sect. 4.1).
This supports the idea that in NGC~7319, as in other 
interacting galaxies, the 
HI was more easily stripped whereas CO stayed
within the optical disk.

In SQ~B  the molecular gas is found in close association 
with the atomic gas and with indicators
of star formation, as described in detail in the previous sections. 
Altogether, this spatial and (in the case of HI also kinematical) coincidence
suggests that we are seeing in-situ molecular 
gas formation possibly provoked by compression of the atomic gas, 
and the subsequent star formation traced by
\halpha \ emission and other tracers.
{  Further support for this comes from the distribution of 
\mhtwo/\mhi. If the molecular gas had been stripped
together with the atomic gas, the molecular-to-atomic gas mass 
ratio would have been
preserved, if no molecular gas formation or 
destruction  took place after the stripping. In this case it would be
an unlikely coincidence that we find the highest value of \mhtwo/\mhi \
close to the HI peak. 
}

\subsection{Tidal Dwarf Galaxies in Stephan's Quintet?}

We observed very different regions in SQ
and our CO observations have shown that the properties
of the ISM are
very diverse. Some of our observed regions have been
claimed to be TDG candidates \citep[e.g.][]{Mendes01,Iglesias01}.
\citet{Duc00} defined a TDG as a {\it self-gravitating} entity
formed out of tidal debris. Although   it is, without  a detailed
kinematical study, difficult to decide  whether a given TDG
candidate is indeed self-gravitating, \citet{Braine00} have argued
that the detection of molecular gas forming in situ gives strong
support to the hypothesis of self-gravitation. Most
objects observed in SQ are, however, different in one way or another 
from the sample of TDGs observed by \citet{Braine00,Braine01}.

SQ~B is the object that most closely resembles TDGs.
Its has a clear tidal origin and we have 
summarized evidence that the molecular gas is forming in situ.
The ratio of molecular gas mass to  the 
\halpha \, luminosity is normal for star forming galaxies.
The star formation rates that can be derived from different tracers
such as \halpha \, emission, 15 $\mu$m emission  
\citep[following the estimate in][]{Xu99} or radio continuum 
(applying eq. 21
in \citet{Condon92}, extrapolated with a Salpeter initial
mass function down to stars with masses above 
0.1 \msun) 
give similar values of about 0.2 \msun yr$^{-1}$. 
{  
These values are in the same range as those found for
TDGs ($0.01 - 0.3$ \msun yr$^{-1}$) and several nearby dwarf galaxies
($0.05 - 0.9$ \msun yr$^{-1}$) \citep{Braine01}
showing that the observations of SQ~B are consistent with the formation of  
a dwarf galaxy in the tidal tail.
}
However, the large extent of the molecular gas covering the
entire tip of the optical tail is surprising and different from
other TDGs. One reason is probably the fact that star formation 
happens at two positions 
along the optical tidal tail: apart from  SQ~B there is a 
weaker star-forming
knot  showing a blue colour and 
\halpha \ emission at the very end of the optical 
tidal tail (at $\alpha_{2000}= 22^h 36^m 12.2^s$ and 
$\delta_{2000} = 33^\circ 57^\prime 30$\arcsec). 
Molecular gas was detected at this position.
Higher resolution observations are necessary in order to find out
whether the molecular gas is concentrated towards the
two  star forming
regions as would be expected if TDGs are forming.

The situation in SQ~A is entirely different.
Whereas the star formation is very  concentrated
at the central position of SQ~A, 
the distribution of the molecular gas is much more extended 
with no maximum at the starburst. This lack of association of 
the molecular gas with the starburst region,
together with the unusual origin of the ISM in this
region brings up strong
doubts whether SQ~A is a TDG, i.e. an object that 
is bound by gravitation.
{   
These doubts are reinforced by the large dynamical mass 
that is deduced from the CO velocity gradient in SQ~A
(see Sect. 3.2). 
Under the assumption that the region around SQ~A is gravitationally bound
and that the velocity gradient is due to 
rotation we can calculate the dynamical mass 
($M_{\rm dyn} = R \triangle V^2/G$,
where $R$ is the radius of the region, $\triangle V$ is 
the velocity gradient along $R$ and
$G$ is the gravitational constant). With $R=30$\arcsec \ and
$\triangle V=80$ \kms \ we derive 
$M_{\rm dyn}= 1.9 \times 10^{10}$ \msun. 
This is much larger 
than the mass  of the molecular and atomic gas at this velocity
($3.8 \times 10^{9}$ \msun). The stellar mass in the starburst
region has been estimated by \citet{Xu99} from a K-band image
to be  $0.8 - 1.6 \times 10^{7}$ \msun , 
much smaller than the gas mass. 
Thus, the dynamical mass is much higher than
the visible mass 
%
%
meaning that this region would need a high fraction of dark matter
in order to be gravitationally
bound. 
This is in contrast to what has been found for a sample of
TDGs \citep{Braine01} where the visible and dynamical masses
were found to be very close obviating the need for
dark matter. We interpret the high dynamical mass
as evidence that this region is {\it not} gravitationally bound
on the size scales observable at the current resolution.
}
The extended  and homogeneous distribution of the molecular gas 
in this region makes it unlikely that a gravitational 
collapse is responsible for its formation.
{   
The star formation in the starburst region SQ~A has rather been
triggered by local processes,
as e.g. by the  pressure of the hot surrounding
gas, as suggested by \citep{Xu99}.
}

Most of the objects in SQ~C possess a gradient in the \halpha \,
emission indicative of rotation \citep{Mendes01} and might 
therefore be 
self-gravitating objects. However, the lack of gas, both
atomic (except for C16/17 and C18/19 the atomic
gas densities are $\le 10^{20}$ \cm) 
as well as molecular, makes them different from 
other TDGs. 
For this reason it is unlikely that they will become 
like TDGs that resemble
dwarf irregulars or blue compact dwarfs as 
they do not have the gas reservoir to continue forming stars
for a long time.

\section{Conclusions}

We have observed CO in several regions in SQ. Our main results are:

\begin{enumerate}

\item In both SQ~A and SQ~B we detected large amounts of molecular
gas of $3.1 \times 10^9$ \msun\ and  $7.0 \times 10^8$ \msun,
respectively. The molecular-to-atomic gas mass ratios are 
high (1.2, respectively 0.5), much larger than what has been  found 
in typical TDGs \citep{Braine01}. The molecular gas is very extended,
over regions of about 15 -- 25 kpc.
            
\item The CO
spectrum in SQ~A consists of lines at  different velocities,
centered at 6030 and 6700 \kms, with most of the emission coming
from the low velocity component,
whereas in SQ~B the CO is
at 6625 \kms. 
The kinematical agreement (velocity, width and spectral
shape) between the CO and HI lines is
very good in both regions, except for a small offset 
in the low-velocity components of the CO and HI lines in SQ~A
(the HI line peaks at 6000 \kms).

\item In SQ~A, the CO emission at 6000 \kms \  is found 
south of the starburst region, covering the southern
part of the HI distribution at this velocity and coinciding with X--ray
and radio continuum emission which are indicative of shocks.
The much weaker CO emission at 6700 \kms \ is more spatially
concentrated.
We conclude that the presence of large amounts of extended
molecular gas at 6000 \kms \ implies that the shocks traced by
X--ray and radio continuum emission only affect the gas at 
6700 \kms \ and that the apparent spatial coincidence is a projection.

\item In SQ~B there is good spatial agreement between the
tracers of star formation (\halpha, 15 $\mu$m and radio continuum)
and the CO emission. The CO peak is slightly offset from
the HI peak towards
a steep HI gradient. We interpret this as indicating that the
molecular gas is forming in-situ, possibly in a region of compressed HI and
that the star formation is taking place subsequently.
This, together with the fact that the emission at different 
wavelengths is similar to that of normal star forming 
galaxies and that SQ~B has
a clear tidal origin, makes it the object in SQ that most resembles a 
TDG. 

\item We have searched without success for CO in several objects around
NGC~7318a/b in which \citet{Mendes01} found \halpha \ emission.
The upper limits for the ratios between molecular gas mass and
SFR (derived from the \halpha \ emission) are  lower than
those found in spiral galaxies and TDGs.
The molecular-to-atomic gas mass ratio is also low, though
not unusual. The low atomic and molecular gas content
of these regions makes them different from TDGs and
will not allow them to sustain star formation for a long time. 

\end{enumerate}

\begin{acknowledgement}
We would like to thank J. Iglesias-P\'aramo and J. V\'\i lchez 
for making available their
\halpha \ data to us, L. Verdes-Montenegro for the HI data,
C. Xu for the ISOCAM and \halpha \,data and S. Gallagher for 
useful information about the star clusters in SQ.
Thanks are also due to the referee, G. Petitpas, for the careful
reading of the manuscript and detailed suggestions for improvements.
VC would like to acknowledge the partial support of JPL contract
960803 and EB acknowledges financial support from CONACyT via project
27607-E. We made use of data from the Canadian Astronomy Data Center,
which is operated by the Dominion Astrophysical Observatory for the National
Research Council of Canada's Herzberg Institute of Astrophysics.

\end{acknowledgement}

{}


\end{document}